# Magnetism in Cr-doped ZnS: Density-functional theory studies

**Xingtao Jia[1,2], Minghui Qin[3] and Wei Yang[4]**
[1] Department of Physics, Beijing Normal University, Beijing, 100875, China
[2] College of Chemistry and Chemical Engineering, China University of Petroleum, Dongying, 257061, China
[3] National Laboratory of Solid State Microstructures and Department of Physics, Nanjing University, Nanjing 210093, China
[4] College of Information, Shanghai Ocean University, Shanghai, 200090, China

E-mail: yangwei@hdpu.edu.cn

**Abstract**
We investigated the magnetism and aggregation trends in cubic $Zn_{1-x}Cr_xS$ using the density-functional theory calculations. We demonstrate that all studied configurations show ground state half-metallic ferromagnetism (HMF); and Cr impurities are energetically favorable to planar cluster into delta-doping structures. The single-layer delta-doping structures of $Zn_{0.75}Cr_{0.25}S$ and $Zn_{0.875}Cr_{0.125}S$ show ferromagnetic stabilization energies ($\Delta E_{AF}$) of 0.551 and 0.561 eV/Cr-Cr pair, respectively. The half-layer delta-doping structure of $Zn_{0.875}Cr_{0.125}S$ and double-layer delta-doping structure of $Zn_{0.75}Cr_{0.25}S$ show $\Delta E_{AF}$ of 0.394 and 0.166 eV/Cr-Cr pair, respectively. Furthermore, our studies indicate that the cubic ZnS/CrS heterostructure, one extreme situation of the delta-doping structure, also shows ground state HMF. The origin of HMF is discussed using a simple crystal field model. Finally, we anticipate the potential spintronics application of $Zn_{1-x}Cr_xS$.

## 1. Introduction

Zn-chalcogenides (ZnX, X= S, Se, Te) based magnetic semiconductors (MSs) have attracted much attention recently for the potential room temperature spintronics application [1-19]. Among them, ZnS based MSs are more promising to integrate into the mainstream silicon technology for similar crystal structure and close lattice parameters. However, although lots of studies have demonstrated ferromagnetism with Curie temperature ($T_C$) above room temperature in ZnX based MSs, the results are always under controversy [13-17]. For MSs, the largest obstacle to the practical spintronics application is the intrinsic aggregation trend of magnetic impurities [20-25]. How to control the clustering of the magnetic impurities always is the hot issue to MSs. Interestingly, some studies show that the magnetic clusters would endow matrix semiconductors with high $T_C$ and huge magnetic moment [23-25]. But the problem is how to pattern the magnetic clusters. The orderly patterned magnetic impurity atoms or clusters are the key to applicable MSs.

Recently, Qian *et al* [26] and Nazmul *et al* [27] introduced the delta-doping scheme into Mn doped silicon and GaAs, respectively. They found that delta-doping would enhance the magnetism. Moreover, theoretical studies also predicted enhanced magnetism in delta-doping structures of (Mn, III)-V semiconductors [28]. Delta-doping is a kind of heterogeneous doping with one more or less impurities layers in a specific face of the matrix semiconductor, which also can be considered as a kind of planar clustering. A prominent advantage of the delta-doping is the facility to achieve higher concentration impurities, which is easy to be realized via the metastable molecular beam epitaxy (MBE) synthesis. According to the Zener model [29], the higher TM impurities doping concentration suggests the higher Curie temperature ($T_C$), which is important to the practical above room temperature applications. So, we can anticipate higher $T_C$ in MSs via the delta-doping scheme.

Compared with the delicate experimental method, the computational methods especially the parameter-free first-principles calculations are widely used to explore the MSs with different doping structures from the ideal homogeneous configurations to extreme inhomogeneous ones such as delta-doping. Generally, density-functional theory (DFT) within local-density approximation (LDA) or generalized-gradient approximation (GGA) tend to underestimate the band gap of the semiconductors and fail to describe the Mott insulator. A mean-field Hubbard potential approximations (U) correction would improve the electronic structure and optical properties of these materials greatly. For Zn-chalcogenides, extensive studies show that the DFT within the LDA would underestimate the band gap, while the GGA would give relatively good descriptions as the LDA+U approach [30-32]. Comparatively, the LDA+U approach would improve the description of the band gap of ZnO greatly for stronger local correlations [30-32]. Correspondingly, the addition of Hubbard term would improve the electronic structure and magnetism of Zn-chalcogenides based MSs by increasing the intermediate bandwidth and stabilizing the ferromagnetic spin order[33], but the effect is not so rudimentarily [34,35], when compared with the ZnO based MSs [36,37].

Here, we report the magnetism in cubic $Zn_{1-x}Cr_xS$ using the DFT calculations. We find that all studied configurations of $Zn_{1-x}Cr_xS$ show ground state half-metallic ferromagnetism (HMF); and Cr impurities are prone to planar clustering into delta-doping structures with enhanced HMF. In the end, we discuss the origination of magnetism and anticipate the potential spintronics application of $Zn_{1-x}Cr_xS$.

## 2. Computational details

We performed spin-polarized total energy DFT calculations in the Perdew-Burke-Ernzerhof GGA [38] using the plane-wave ultrasoft pseudopotential method [39] as implemented in the Cambridge serial total energy package (CASTEP) code. [40] To study the effect of doping configurations on the magnetism, a 1x1x2 conventional supercell as shown in figure. 1(a) was used for cubic $Zn_{0.75}Cr_{0.25}S$. Therein, N110 is a kind of single-layer delta-doping in cubic ZnS (001). There, we find that Cr impurities are prone to planar cluster into delta-doping structure. To study the aggregation trend of magnetic impurities in delta-doping configuration under lower doping concentration, a cell containing 32 atoms corresponding to $Zn_{0.875}Cr_{0.125}S$ with lattice parameters a= b= 7.65 Å and c=10.82 Å, as shown in figure. 1(b), was used. Furthermore, a cell containing two CrS layers along [001] direction as shown in figure 1(c) were used to study the double-layer delta-doping structure, and a cubic ZnS/CrS superlattice containing 4 layers of ZnS and 4 layers of CrS as shown in figure. 1(d) was used to simulate a kind of extreme situation of delta-doping. Typically, the effect of the cutoff energies and Monkhorst-Pack grids on the ferromagnetic stabilization energy (energy difference between antiferromagnetic and ferromagnetic states, $\Delta E_{AF}= E_{AFM}- E_{FM}$) of the delta-doping structure of $Zn_{0.75}Cr_{0.25}S$ (N110 in figure 1(a)) is checked by using different cutoff energies (350, 450 and 550 eV) and Monkhorst-Pack grids (6x6x3, 8x8x2 and 8x8x4), the result shows that the difference of $\Delta E_{AF}$ is less than 0.002 eV. To save the calculation expense, the lowest energy cutoff and Monkhorst-Pack grid are used in the calculations. For $Zn_{0.75}Cr_{0.25}S$ in figure 1(a) and $Zn_{0.875}Cr_{0.125}S$ in figure 1(b), the Monkhorst-Pack grids of 6x6x3 and 4x4x3 were used, respectively; for the double-layer delta-doping structure of $Zn_{0.75}Cr_{0.25}S$ [figure 1(c)] and ZnS/CrS heterostructure [figure 1(d)], the Monkhorst-Pack grids were set 6x6x2. Both the lattice parameters and atomic positions were fully relaxed in all calculations. Otherwise, the magnetic moments keep variable to find out the lowest energy state. Through the letter, the formation energy of Cr substitution impurities ($Cr_{Zn}$) in ZnS was calculated with the well-established formula [41,42], which is defined as

$$E_f[Cr_{Zn}] = E[Cr_{Zn}] - E[bulk] - n\mu_{Cr} + n\mu_{Zn} \qquad (1)$$

where $E[Cr_{Zn}]$ and $E[bulk]$ are the total energies of $Zn_{1-x}Cr_xS$ and pure ZnS reference structures (as calculated with the same size supercell), respectively. Here, $\mu_{Cr}$ and $\mu_{Zn}$ are the atom chemical potentials, which represent the energy of the reservoirs with which atoms are being exchanged. The chemical potential depends on the experimental conditions under which the materials are fabricated. Here, we just considered the moderate condition:

$$\mu_{Zn} = \mu_{Zn}(bulk) - 0.5\Delta H(ZnS) \qquad (2)$$

$$\mu_{Cr} = \mu_{Cr}(bulk) - 0.5\Delta H(CrS) \qquad (3)$$

In these equations, $\Delta H(ZnS)$ and $\Delta H(CrS)$ are the heat of formation of cubic ZnS and CrS, respectively, $\mu_{Zn}$ (bulk) and $\mu_{Cr}$ (bulk) are calculated from antiferromagnetic cubic chromium (Cr) and hexagonal zinc (Zn), respectively. Generally, $Cr_{Zn}$ formation energy shows positive value, the lower the $Cr_{Zn}$ formation energy, the easier the formation of $Cr_{Zn}$ impurities.

## 3. Results and discussion

The calculated ferromagnetic stabilization energy $\Delta E_{AF}$, $Cr_{Zn}$ formation energy, total and Cr sites magnetic moments of $Zn_{1-x}Cr_xS$ with different doping configurations are given in table 1. Apparently, all calculated doping configurations show positive $\Delta E_{AF}$ (positive $\Delta E_{AF}$ means that the FM is favored over AFM, and the larger the $\Delta E_{AF}$, the stabler the FM) and integral total magnetic moment, indicating ground state HMF (The paramagnetic states are also calculated for $Zn_{1-x}Cr_xS$. The results show that the total energies of the paramagnetic states are substantially higher than the ferromagnetic and antiferromagnetic states). In addition, many of them show robust HMF with large $\Delta E_{AF}$ close to traditional magnetic metals and alloys. For $Zn_{0.75}Cr_{0.25}S$ shown in figure 1(a), the single-layer delta-doping (N110) and Cr dimer (N011) structures show larger $\Delta E_{AF}$ and lower $Cr_{Zn}$ formation energy than the homogeneous-doping structure (N002) and N112, indicating a clustering trend of the Cr impurities. Among them, N110 shows largest $\Delta E_{AF}$ (0.551 eV/Cr-Cr pair) and lowest $Cr_{Zn}$ formation energy (1.127 eV), indicates that the delta-doping structure is not only most magnetically stable but also energetically favorable. That is, Cr impurities are energetically favorable to planar cluster into delta-doping structure. Comparatively, the delta-doping structures of cubic (Ga, Mn)As [27] and (Ga, Mn)N [43], which have been experimentally realized, show larger $Ga_{Mn}$ formation energy of 1.579 and 1.738 eV (under moderate conditions), respectively, when the doping concentration is 25%. Deductively, the delta-doping structure of (Zn, Cr)S is more feasible to experimental fabricate comparatively. To explore the planar clustering trend in the delta-doping structure under lower doping concentration, we calculated some doping structures of $Zn_{0.875}Cr_{0.125}S$ as shown in figure 1(b). Apparently, N111 (Cr dimer) and N200 (a kind of heterogeneous half-layer delta-doping) show larger $\Delta E_{AF}$ and lower $Cr_{Zn}$ formation energy than N220 (homogenous half-layer delta-doping) and N222 (homogeneous doping). Among them, N200 shows largest $\Delta E_{AF}$ (0.394 eV/Cr-Cr pair) and N111 shows lowest $Cr_{Zn}$ formation energy (1.229 eV). That is, Cr impurities trend to clustering into Cr dimer under lower doping concentration. Comparatively, we investigated the single-layer delta-doping structure of $Zn_{0.875}Cr_{0.125}S$, which shows $\Delta E_{AF}$ of 0.561 eV/Cr-Cr pair and $Cr_{Zn}$ formation energy of 1.105 eV. Apparently, the single-layer delta-doping structure shows larger $\Delta E_{AF}$ and lower $Cr_{Zn}$ formation energy than the half-layer delta-doping and Cr dimer structures with same doping concentration, and the lower concentration delta-doping structure shows larger $\Delta E_{AF}$ and lower $Cr_{Zn}$ formation energy than the higher one. Furthermore, we investigated the double-layer delta-doping structure of $Zn_{0.75}Cr_{0.25}S$ and ZnS/CrS heterostructure. The results indicate ground state HMF in both samples with $\Delta E_{AF}$ of 0.166 and 0.055 eV/Cr-Cr pair and $Cr_{Zn}$ formation energy of 1.014 and 0.980 eV, respectively. (Here, we define two AFM states for the ZnS/CrS heterostructure, one is defined by alternating planes spin up/spin down in the direction [001] of cubic ZnS, another is defined by spin up and down in the same plane perpendicular to [001]. The results show that the former is more energetically favorable.) The latter shows smaller $\Delta E_{AF}$ which is closer to the cubic CrS (0.063 eV/Cr-Cr pair). Inspecting the relations between the delta-doped CrS layers and $\Delta E_{AF}$ and $Cr_{Zn}$ formation energy, we can see that both the $\Delta E_{AF}$ and $Cr_{Zn}$ formation energy show a trend to decrease as the delta-doped CrS layers increasing, which decrease fast at first and get slower when the delta-doped CrS go beyond two layers. Observing the total and Cr sites magnetic moments, we can see the doping configurations show no effect on the total magnetic moments, while little on the Cr sites magnetic moments. Generally, the more homogeneous structures shows lower Cr sites magnetic moments, larger Cr-Cr distance, smaller $\Delta E_{AF}$, and larger $Cr_{Zn}$ formation energy. The Cr dimer and single-layer delta-doping structures show larger Cr sites magnetic moments, smaller Cr-Cr distance (compared with the Zn-Zn distance of 3.828 Å in ZnS), larger $\Delta E_{AF}$, and smaller $Cr_{Zn}$ formation energy. The double-layer delta-doping and ZnS/CrS heterostructure show larger Cr sites magnetic moments, larger Cr-Cr distance (which is larger than the Zn-Zn distance in ZnS, but smaller than the Cr-Cr distance in cubic CrS), smaller $\Delta E_{AF}$, and smaller $Cr_{Zn}$ formation energy. Apparently, the $\Delta E_{AF}$ shows a clear Cr-Cr distance and structures dependence, and $Cr_{Zn}$ formation energy shows a aggregation dependence. The smaller Cr-Cr distance structures always show larger $\Delta E_{AF}$, and intense aggregation structures show lower $Cr_{Zn}$ formation energy. In summary, all studied doping configurations of $Zn_{1-x}Cr_xS$ show ground state HMF, and Cr impurities are prone to planar cluster into delta-doping structures with enhanced HMF.

Figure 2 shows the spin- and sites-resolved density of states (DOS) of the homogeneous-doping and single-layer delta-doping structures of $Zn_{0.75}Cr_{0.25}S$ in their ferromagnetic states. Apparently, both samples show pronounced majority HMF around the Fermi level ($E_F$), and the DOS curves of the homogeneous-doping structure are more regular than the delta-doping one, which should be attributed to the enhanced Jahn-Teller effect in the latter. The homogeneous-doping structure shows S-Cr-S angle of 109.47$^O$ and Cr-S distance of 2.385 Å (larger than the Zn-S distance of 2.344 Å in ZnS), while the single-layer delta-doping structure shows two S-Cr-S angles of 105.66$^O$ and 111.41$^O$ and Cr-S distance of 2.389 Å. The existence of Cr impurities expand the ZnS substrate isotropically in the former, while distort the ZnS substrate anisotropically in the latter.

Apparently, the Jahn-Teller effect is more remarkable in the delta-doping structure. For both doping structures, both Cr and S atoms contribute to the majority HMF around the $E_F$, evidencing strong hybridization between Cr and S atoms. The impurities DOS peaks in the band gap of pure ZnS can be associated with the $e$-$t_2$ splitting expected from a simple crystal field model and the spontaneous Jahn-Teller distortion of the bonding in cubic ZnS. Hordequin *et al* [44] and Qian *et al* [26] have investigated the relation between the electronic structure and the spin-flip transitions (electronic phase transition from half-metal to normal ferromagnet or semiconductor) temperature $T^*$. They found that the spin-flip gap $\delta$ is relevant to the $T^*$, the wider the $\delta$, the higher the $T^*$. Inspecting the DOS of the homogeneous- and delta-doping structures of $Zn_{0.75}Cr_{0.25}S$ around the $E_F$, we can see that the latter shows wider gap $\delta$ than the former. So, just in view of the electronic structure, we can deduce that the delta-doping structure should possess higher $T^*$ than the homogeneous-doping structure. Actually, besides electronic structure, the ferromagnetic stabilization energy $\Delta E_{AF}$ shows impact on the spin-flip transition temperature $T^*$ too. Basically, the larger the $\Delta E_{AF}$, the higher the $T^*$. Inspecting the table 1, we can see that the delta-doping structures of $Zn_{1-x}Cr_xS$ show larger $\Delta E_{AF}$ than the homogeneous-doping ones. So, just in view of the $\Delta E_{AF}$, we can see that the delta-doping structure should possess higher $T^*$ than the homogeneous-doping one. In summary, the single-layer delta-doping structure shows enhanced HMF than the homogeneous-doping one in both electronic structure and ferromagnetic stability.

Figure 3 shows DOS of the double-layer delta-doping structure of $Zn_{0.75}Cr_{0.25}S$ and ZnS/CrS heterostructure in their ferromagnetic states. Apparently, both samples show pronounced majority HMF around the $E_F$. Compared with the single-layer delta-doping structure (figure 2), the DOS of the double-layer delta-doping structure of $Zn_{0.75}Cr_{0.25}S$ and ZnS/CrS heterostructure, especially the impurity DOS, broaden significantly with more fingerprint peaks, which should be attributed to the enhanced Jahn-Teller effect here. Moreover, another difference would be the spin-flip gap $\delta$. For the single-layer delta-doping structure, the gap $\delta$ is defined by the $E_F$ and the impurity band maximum in the majority-spin channel; while defined by the $E_F$ and the conduction-band minimum (CBM) of the double-layer delta-doping structure of $Zn_{0.75}Cr_{0.25}S$ and ZnS/CrS heterostructure in the minority-spin channel.

The origin of HMF in $Zn_{1-x}Cr_xS$ can be depicted by a simple crystal field model, where one Cr atom lies in the tetrahedral center of four S atoms. Figure 4 schematic describes the $p$-$d$ hybridization of Cr in S tetragonal ($T_d$) field. Therein, the Cr $d$ orbitals would experience an $e$-$t_2$ splitting, which would hybridize with the neighbor S $p$ orbitals further and then lead to HMF. For symmetric reason, the Cr $t_2$ orbitals hybridize with S $p$ strongly while Cr $e$ is weak. The energy difference between the spin-up and spin-down channel is due to the spin exchange splitting. Apparently, in Cr doped ZnS, the spin exchange splitting energy is larger than crystal field splitting energy. That is, the crystal field splitting energy is smaller than the electron pairing energy. So, the system would follow a high-spin arrangement. To lower the total energy, the system would experience a spontaneous Jahn-Teller distortion. Where, only the lower degenerate energy levels are partially filled. From figure 4, we can see the majority states around the $E_F$ are mainly from the bonding $p$-$d$ hybrids orbitals, and the half-metallic gap is form by the energy difference between the minority bonding $p$-$d$ hybrids orbitals and Cr $e$ orbitals. So, any factor strengthened the $p$-$d$ hybridization ($e$-$t_2$ splitting) would enhance the HMF. Delta-doping is a kind of controllable distortion, which can enhance the spontaneous Jahn-Teller effect by breaking the symmetry further. Consequently, the total energy of the system is further lowered.

For $Zn_{1-x}Cr_xS$, the ground state magnetism is determined by the competition between the ferromagnetic double-exchange interaction and anti-ferromagnetic super-exchange interaction, and the former demonstrated to be dominating [1,9]. Inspecting the table 1, it is easy to see that the nearer the Cr-Cr distance, the stabler the FM. Obviously, both the ferromagnetic and anti-ferromagnetic coupling in $Zn_{1-x}Cr_xS$ are getting stronger, and the former is predominant over the latter as the Cr-Cr distance gets close. According to Sato and Katayama-Yoshida [1], the itinerant properties of the spin-polarized conduction electrons play an important role in the stabilization of the ferromagnetic state in ZnX based MSs. Apparently, compared with doping configurations containing two Cr impurities, the single-layer delta-doping structures of the cubic $Zn_{1-x}Cr_xS$, provide more direct (effect) channel for valance electrons (Cr $t_2$) to itinerate. Consequently, the ferromagnetic stability of the single-layer delta-doping structures is substantially enhanced. Comparatively, the double- and multi-layer delta-doping structures show decreasing $\Delta E_{AF}$ (approaching to the value of cubic CrS) as the delta-doping layers get thicker. Here, the ferromagnetism can be depicted by the simple crystal field model [45].

Typically, we studied the sites-resolved magnetic moments of the single-layer delta-doping structure of $Zn_{0.75}Cr_{0.25}S$. Therein, only Cr and first-neighbor S atoms show magnetic moments of 4.30 and 0.15 $\mu_B$, respectively, while the other atoms show zero magnetic moments. Apparently, the delta-doping structure shows a two-dimensional HMF, and the magnetic coupling in Cr doped ZnS has a short-range nature.

Theoretically, we can achieve fractional- or multi-layer delta-doping accurately via the metastable MBE synthesis in the laboratory by modulating the flux ratio of the precursors. Practically, the ideal synthesis is hard to realize, especially mass-production in the factory. There are many factors that determine the distribution of magnetic impurities during MBE synthesis such as the temperature, fluctuation of the precursors flux ratio, surface flatness of the matrix semiconductor, the post treatment temperature and ambience, and so on. That is, the magnetism in the MSs via the MBE synthesis is more dynamics determined rather than energetics. As a result, mostly, the magnetism in MSs is metastable, which depends crucially on the microscopic distributions of the magnetic impurities clusters [23-25,46]. The global magnetism is the collective effect of different magnetic domains. Interestingly, as discussed above, all studied doping configurations of $Zn_{1-x}Cr_xS$ show ground state HMF. This is a valuable character, which would make the metastable MBE synthesis more technologically feasible. Moreover, ZnS not only shows small lattice mismatch with silicon but also shares the similar crystal structure. That is, $Zn_{1-x}Cr_xS$ not only shows ground state HMF, but also has good compatibility with silicon. So, we speculate that $Zn_{1-x}Cr_xS$ may be a more practical spintronics material integrated into the mainstream semiconductor technology.

## 4. Conclusions

In Summary, we have investigated the magnetism and aggregation trends in $Zn_{1-x}Cr_xS$ using the DFT calculations. We find that all studied doping configurations here show ground state HMF, and Cr impurities are prone to plane cluster into the delta-doping structures with enhanced HMF. The single-layer delta-doping structure of $Zn_{0.75}Cr_{0.25}S$ and $Zn_{0.875}Cr_{0.125}S$ show $\Delta E_{AF}$ of 0.551 and 0.561 eV/Cr-Cr pair, respectively, which are substantial higher than that of many compound semiconductors based MSs. Moreover, our studies also demonstrated ground state HMF in the half-, double-, and multi-layer delta-doping structures of $Zn_{1-x}Cr_xS$. The magnetic coupling in $Zn_{1-x}Cr_xS$ shows a short-range nature, and the origin of HMF is discussed using a simple crystal field model. Finally, we anticipate the potential spintronics application of $Zn_{1-x}Cr_xS$.

## Acknowledgments

The work is partially supported by the Postgraduate Innovation Foundation of China University of Petroleum under grant No. B2008-8, and National Laboratory of Solid State Microstructure of Nanjing University (LSSMS) under grant No. M21001.

**Table 1.** The calculated ferromagnetic stabilization energy $\Delta E_{AF}$ (eV), $Cr_{Zn}$ formation energy $E_f$ (eV), and total and Cr sites magnetic moments of $Zn_{1-x}Cr_xS$ with different doping configurations shown in figure 1. Therein $d$ is the nearest distance between Cr-Cr pair, and the total magnetic moment is the sum of all magnetic moments normalized to the unit contained one Cr atom.

| | Magnetic ground state | $d$ (Å) | $\Delta E_{AF}$ (eV) | $E_f$ (eV) | $m_{tot}$ ($\mu_B$) | $m_{Cr}$ ($\mu_B$) |
|---|---|---|---|---|---|---|
| (a) $Zn_{0.75}Cr_{0.25}S$ | | | | | | |
| N112 | HMF | 5.397 | 0.060 | 1.414 | 4.00 | 4.24 |
| N002 | HMF | 5.446 | 0.004 | 1.398 | 4.00 | 4.24 |
| N011 | HMF | 3.679 | 0.407 | 1.197 | 4.00 | 4.26 |
| N110 | HMF | 3.809 | 0.551 | 1.127 | 4.00 | 4.30 |
| Double-layer delta-doping | HMF | 3.833 | 0.166 | 1.014 | 4.00 | 4.32 |
| (b) $Zn_{0.875}Cr_{0.125}S$ | | | | | | |
| N220 | HMF | 5.405 | 0.012 | 1.450 | 4.00 | 4.24 |
| N200 | HMF | 3.806 | 0.394 | 1.324 | 4.00 | 4.26 |
| N111 | HMF | 3.534 | 0.304 | 1.229 | 4.00 | 4.26 |
| N222 | HMF | 7.678 | 0.113 | 1.413 | 4.00 | 4.24 |
| Single-layer delta-doping | HMF | 3.815 | 0.561 | 1.105 | 4.00 | 4.30 |
| (c) $Zn_{0.5}Cr_{0.5}S$ | | | | | | |
| ZnS/CrS heterostructure | HMF | 3.841 | 0.055 | 0.980 | 4.00 | 4.32 |

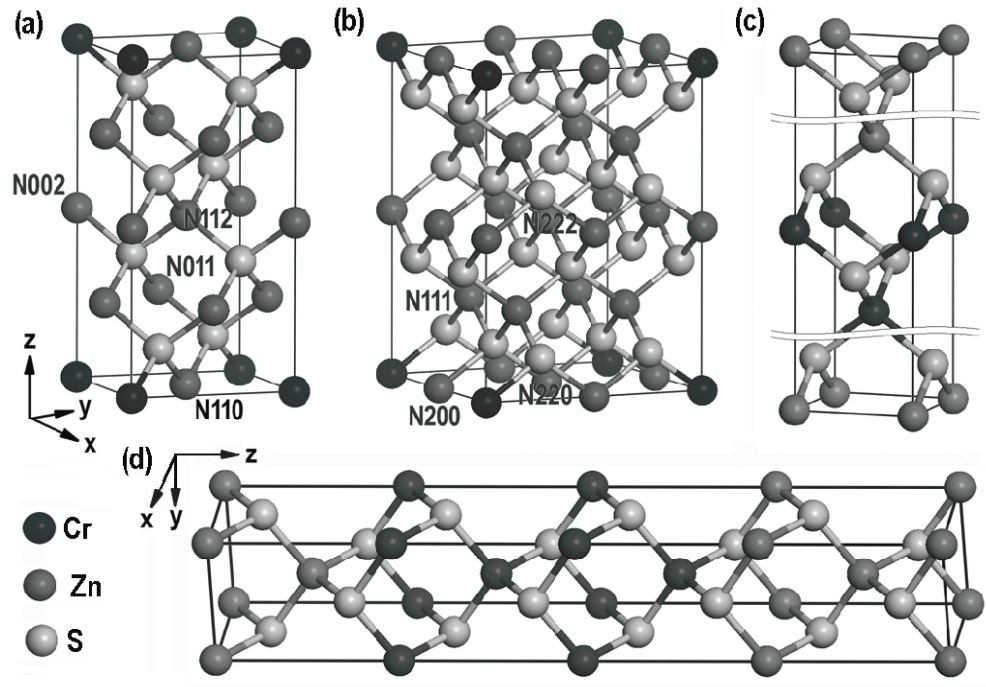

**Figure 1.** Scheme of (a) $Zn_{0.75}Cr_{0.25}S$, (b) $Zn_{0.875}Cr_{0.125}S$, (c) double-layer delta-doping structure of $Zn_{1-x}Cr_xS$, and (d) ZnS/CrS heterostructure. Here, *Nxyz* in (a) and (b) represent configurations containing a Cr-Cr pair with one Cr at the origin and another at *Nxyz* position.

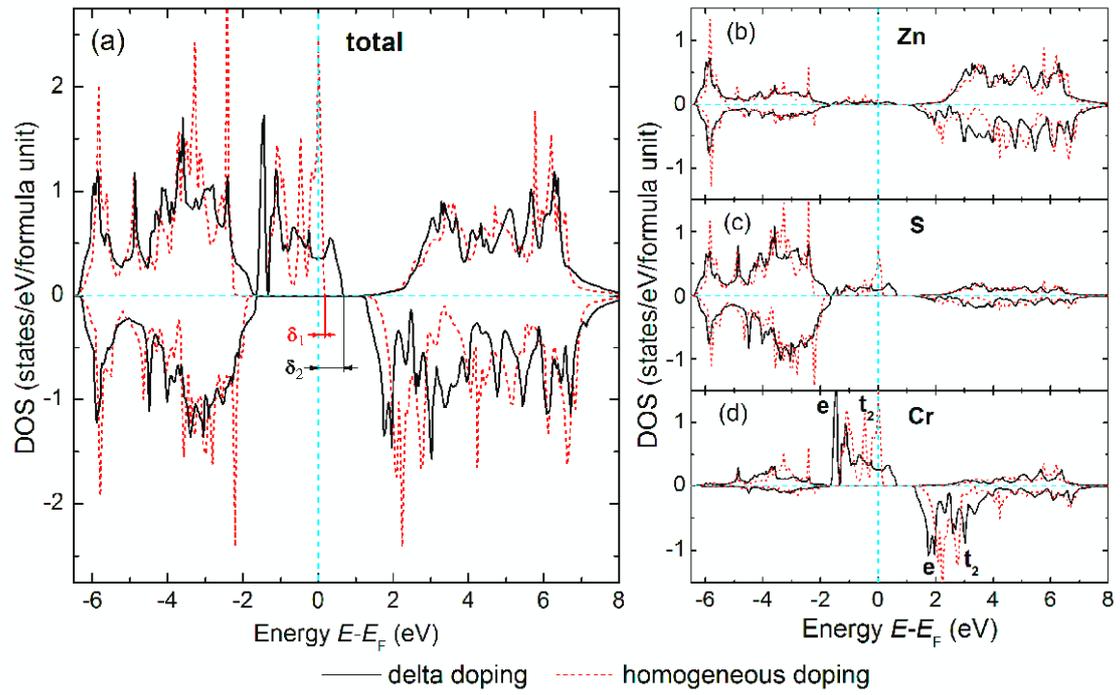

**Figure 2.** Spin- and sites-resolved DOS of the homogeneous-doping and (sngle-layer) delta-doping structures of $Zn_{0.75}Cr_{0.25}S$.

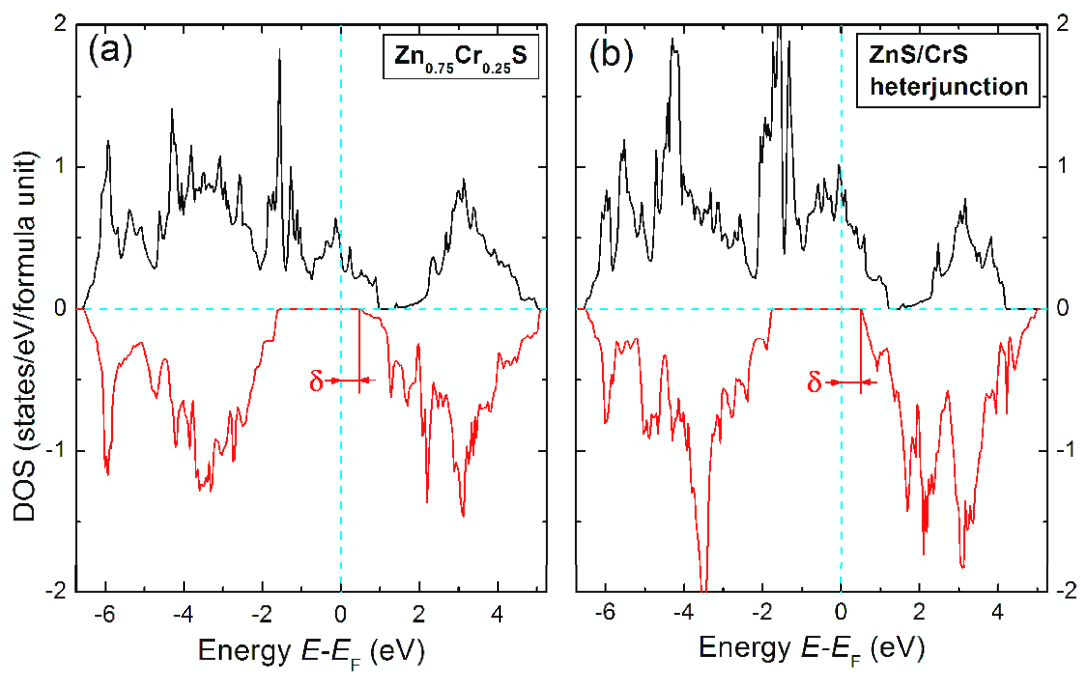

**Figure 3.** Spin-resolved DOS of (a) the double-layer delta-doping structure of $Zn_{0.75}Cr_{0.25}S$ and (b) ZnS/CrS heterostructure

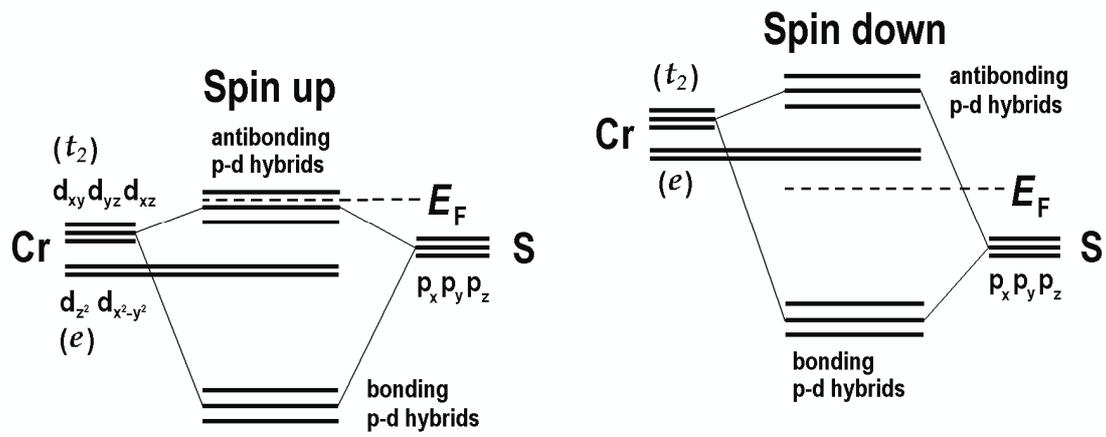

**Figure 4.** The scheme of the *p-d* hybridization in $Zn_{1-x}Cr_xS$.